# Rationale for METI

## Alexander L. Zaitsev, IRE, Russia

**alzaitsev@gmail.com**

Transmission of the information into the Cosmos is treated as one of the pressing needs of an advanced civilization. The inability to meet this need, the forced withdrawal into self-imposed isolation can lead to the extinction of civilization.

### Introduction

For more than 50 years of space exploration Earth space probe ("Voyager 1") was able to fly only 16 light-hours, which is about 2000 times smaller than the distance to the nearest star. Moving with such speed that the probe, which was launched in 1977, will reach this star in a few tens of thousands of years. Therefore, it is obvious that from civilizations such as ours, we can expect only the arrival of electromagnetic signals. The search for such signals is conducted by SETI programs (Search for Extra-Terrestrial Intelligence).

With regard to civilizations that are far ahead of us in their development and whose representatives are able to arrive here, we are not able to hide. Such a highly developed civilization for us – are like something supernatural and mysterious.

They will have long ago found an unusual planet, for more than 200 million years with a spectral line of oxygen, an indicator of life processes. This finding would bring a close eye on us, so that over time they would detect the manifestations of intelligent life. It should be clearly understood, then, they will find intelligent life not by our interstellar radio messages (IRMs) sent into space during realization of METI programs (Messaging to Extra-Terrestrial Intelligence), but primarily by radar sounding signals from Earth's radio telescopes. The probability of detecting radar sounding signals is about a million times higher than the probability of detection of our IRMs (Zaitsev, 2008).

Well-known the key role of radar astrometry in the fast detection of dangerous space objects (Yeomans et al., 1987), and there is no other way for such rapid diagnosis. Therefore, we can not impose a ban on the radiation into space of powerful radar signals. *And it is very important to understand:* both NEO Radar Research and Messaging to ETI are using the same powerful instruments: Arecibo Radar Telescope (ART), Goldstone Solar System Radar (GSSR), and Evpatoria Planetary Radar (EPR). Also, *it is very important to understand:* "addressless" RADAR transmissions and targeted METI are absolutely equivalent, because monster su-per-aggressive and super-powerful ETIs may live anywhere.

Therefore, all the talk about alien invasion and the danger of METI I regard as idle and pseudoscientific. These numerous debates carry out, mainly by those who are far from astronomy and do not realize the size of the Universe. The following is a justification of METI, this new type of human activity, which aims at transforming our civilization into an object of possible detection by extraterrestrial civilizations. Such activities are humane and selfless, it allows us to understand our own way and do not fade away in the future because of "apathy".

## METI as the need of a highly developed civilization

More than 40 years ago, Nicholas Kardashev (1971) expressed the profound idea that the transmission of information into the Cosmos, to the address of alleged "brothers on reason" is a vital and a natural need of highly developed civilization. He wrote: "There are reasons to believe that transmission of information is one of the basic conditions of existence for super-civilizations". It is clear that METI is treated not as a "bait" to attract other civilizations and to ensure the success of terrestrial searching, but as something immeasurably greater, namely, as one of the fundamental requirements of an advanced civilization.

Extremely interesting is the historical aspect of the problem. We give only two examples among many. In the early of 19th century, Carl Gauss was thinking about how to tell the aliens of the existence of intelligent beings on Earth. In 1896 Konstantin Tsiolkovsky published in the weekly "Kaluga Herald" an article with the project on the same topic. The main question related to these and many similar projects (Schirber, 2009), is: "How to understand the interest of the outstanding scientists of the past to this problem? Why do they think about these topics and with which connected such need?" The issue is not as simple as it seems at first glance, it should not be reduced to the appearance of possible eccentricities of those noted scientists…

## METI and the Great Silence

In 1999, after the development and transmission by us from Evpatoria of the first multi-page interstellar radio messages «Cosmic Call» (Zaitsev, 2011), an American classic in the field of radar studies of asteroids and comets Steven Ostro sent us his unpublished work: «Project Moonbeam: An Omnidirectional Radio Beacon for the Lunar Farside. JPL, October 1989 ". In this paper, he proposed to create a powerful beacon for regular interstellar broadcasts. Particularly memorable is a phrase that deserves to be a maxim: "We might conclude that it is better to give than to receive, and that the war on Great Silence must begin at home". The earlier terrestrial planetary consciousness begins to understand and accept this idea, the better!

And while searching, the various SETI programs spent hundreds of times longer than the transmission with METI programs. This paradoxical disparity of effort, a passionate desire to receive and nothing to give, was subsequently called "The SETI Paradox" (Zaitsev, 2006). A trivial consequence of this paradox is an explanation of the Silence of the Universe: "If not only earthly but also other planetary consciousness are so inclined that they prefer to receive rather than give, the search does not make sense, because the Universe is silent".

Another conceivable reason for Silence is intimidation of our society, by scientists and science-fiction writers with the threat of alien invasion. At one time we wrote about this in the article (Zaitsev et al., 2005): «In conclusion, we subscribe to one possible solution to the Fermi Paradox: Suppose each extraterrestrial civilization in the Milky Way has been frightened by its own SETI leaders into believing that sending messages to other stars is just too risky. Then it is possible we live in a galaxy where everyone is listening and no one is speaking. In order to learn of each others' existence - and science - someone has to make the first move».

It is necessary to understand and remember that the transmission of interstellar radio messages from the Earth is filled with meaning and our own searching for radio messages from other civilizations. After all, if ever whip up hysteria of alien invasion, ban those who engage in METI, calling their actions irresponsible and reckless to the point of idiocy, the question arises – whose messages are the SETI Institute and other groups in SETI seeking? Does the acronym "SETI" deserve to be decoded as the Search for Extra-Terrestrial Idiots?

## Isolationism as a possible cause of extinction of civilizations

I do not know definitely, but it seems to me that Sebastian von Hoerner was the first, who in the 1960s indicated that "apathy" or "Loss of Interest" represents real cause of extinction of advanced Civilizations. In the Russian language is the phrase «одиночная камера», which correspond to the word combination "one-man island" in English. I can not speak for all, but I do not want to live in a cocoon, in a "one-man island", without any rights to send a message outside, because such life is not interesting!

Similarly, the prohibition of message transmission converts the Earth into "one-civilization island". I think that it is not interesting for inhabitants to live in such enforced self-isolation, in such lurker-like Civilization! Civilizations, which are forced to hide and tremble because of farfetched fears, is doomed to extinction...

Thus, summing up, we can conclude that the struggle against ONE mythical ET-threat by means of prohibition of any Radar Astronomy transmission and any sending messages to ETIs, creates TWO real problems: defencelessness in the face of Asteroid Hazard and the threat of very probable extinction of such self-isolated civilization due to "apathy".